\documentclass[%
 reprint,
 amsmath,amssymb,
 aps,
]{revtex4-1}

\usepackage{graphicx}
\usepackage{dcolumn}
\usepackage{bm}
\usepackage{color}
\usepackage{float} 

\begin{document}

\preprint{APS/123-QED}

\title{ Spatially encoded light for Large-alphabet Quantum Key Distribution }

\author{T.B.H. Tentrup}
\author{W.M. Luiten}
\author{R. van der Meer}
\author{P. Hooijschuur}
\author{P.W.H. Pinkse}
 \email{p.w.h.pinkse@utwente.nl}
\affiliation{Complex Photonic Systems (COPS), MESA+ Institute for Nanotechnology,
University of Twente, P.O. Box 217, 7500 AE Enschede, The Netherlands}

\date{\today}

\begin{abstract}
 Most Quantum Key Distribution protocols using a two-dimensional basis such as HV polarization as first proposed by Bennett and Brassard in 1984, are limited to a key generation density of 1 bit per photon. We increase this key density by encoding information in the transverse spatial displacement of the used photons. Employing this higher-dimensional Hilbert space together with modern single-photon-detecting cameras, we demonstrate a proof-of-principle large-alphabet Quantum Key Distribution experiment with $1024$ symbols and a shared information between sender and receiver of $7$~bit per photon.
\end{abstract}

\pacs{Valid PACS appear here}
\maketitle


\section{\label{sec:Introduction}Introduction}
Human society relies increasingly on the availability of affordable and high speed communication, which fosters the need of high key-rate generating cryptography. Recent progress in the development of quantum computers \cite{barends2016digitized,brecht2016multilayer,aasen2016milestones,saffman2016quantum,wang201816} threatens the widely used cryptographic methods, which rely on computational assumptions \cite{shor1994algorithms,menezes1996handbook}. A possible solution is Quantum Key Distribution (QKD) of which the security is only based on quantum physics and not on any computational assumption. The first QKD protocol BB84 \cite{bennet1984quantum} uses the two-dimensional polarization basis to encode information in photons. Therefore, the alphabet is limited to two symbols, "0" and "1", with a maximum information content of 1~bit per photon. Since the generated key is used as a one-time pad, this is a bottleneck especially for encrypted video communication \cite{liao2017satellite}.
\par
There are two approaches to increase the key generation rate. One is to increase the repetition rates of photon generation \cite{yuan2016directly} and detection \cite{patel2014quantum}, which is inherently limited by dead times and jitter of the detectors \cite{brougham2016information}. The other approach is to exploit properties of photons besides the polarization to increase the dimensionality of the Hilbert space \cite{bechmann2000quantumprl,bechmann2000quantum}. A higher dimensional Hilbert space leads to a higher information content of the photons and finally increases the key generation rate. Moreover, the error rates introduced by eavesdropping are larger, resulting in an increased security \cite{cerf2002security,gisin2002quantum,scarani2009security}.
\par
Several methods of high-dimensional QKD have been demonstrated, including time-bin \cite{ali2007large,nunn2013large,zhong2015photon,islam2017provably}, orbital angular-momentum \cite{mafu2013higher,mirhosseini2015high,krenn2014generation,sit2017high} and transverse momentum \cite{walborn2006quantum,etcheverry2013quantum}. Comparing the last two spatial encoding schemes, transverse momentum states have the following advantages. Assuming a realistic sender-receiver configuration with finite-size apertures, a diffraction-limited spot translated in an x,y-plane has a higher capacity limit than the pure OAM states, since they form a subset of Laguerre-Gauss modes \cite{zhao2015capacity,kahn2016twist}. Together with the ease of generating a Fourier-transformed mutually unbiased basis with lens optics, spatial translation states of single photons is a promising candidate for very-high-dimensional QKD.
\par
In principle a scan mirror could be used to spatially translate single photons. However, to correct for disturbances a Spatial Light Modulator (SLM) is more flexible and also allows to use wavefront-shaping methods \cite{vellekoop2007focusing}. The SLM allows to change the phase and amplitude of a wavefront by use of holographic methods.
\par
In this paper we experimentally demonstrate very-high-dimensional QKD with $1024$ distinguishable symbols in two mutually unbiased bases with a shared information of $7$~bit per sifted photon. This value is higher than previously reported values of $2.05$~bit for OAM states \cite{mirhosseini2015high} and comparable to the values demonstrated in time-energy QKD \cite{zhong2015photon}. We give finite-key security arguments for claiming an error-corrected and privacy-amplified secret-key rate of the final key of more than $0.5$~bit per photon.         


\section{\label{sec:Experiment}Experiment}


\begin{figure*}[htb]
\begin{center}
\includegraphics[width=0.8\textwidth]{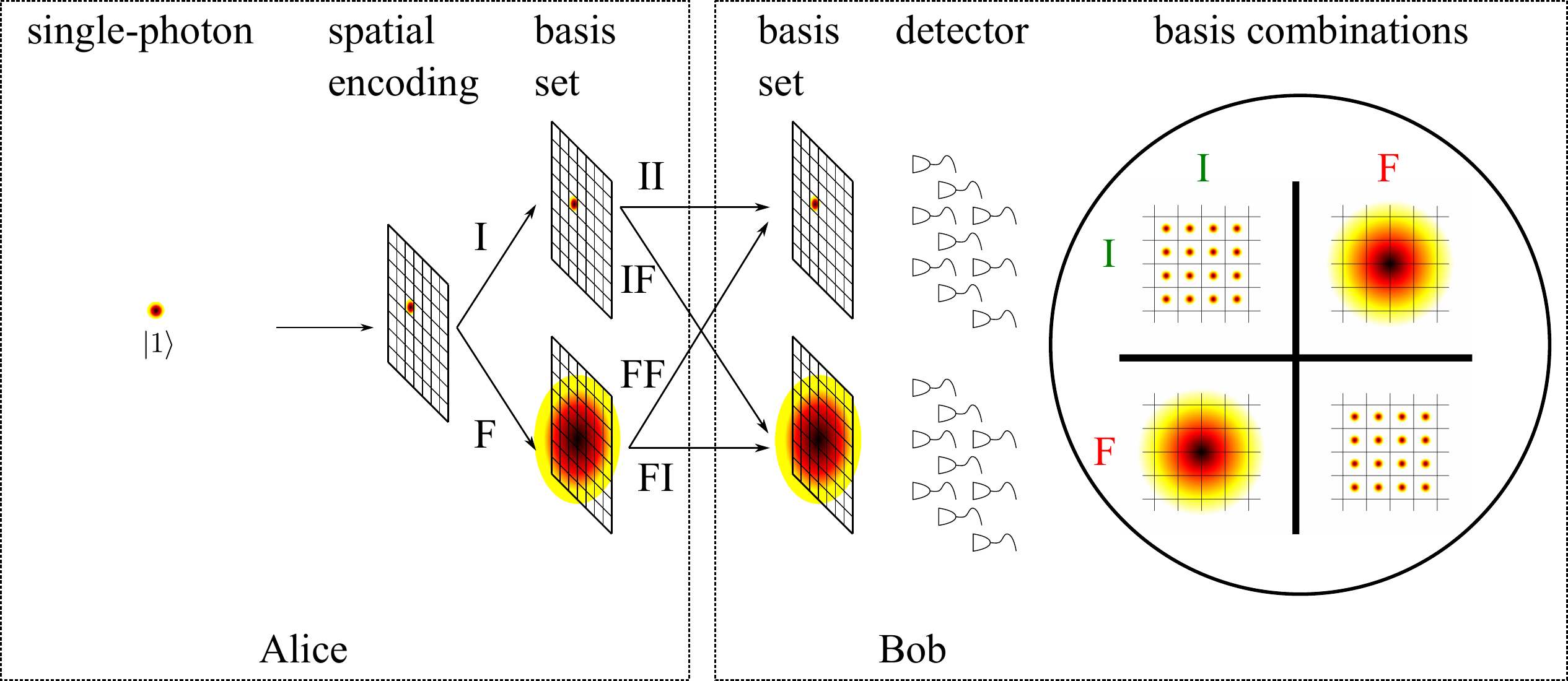}
\caption{Illustration of our spatial encoding and decoding scheme and possible basis choices therein. The single-photon state $\left| 1\right>$ is encoded in the x,y- translation basis formed by shifting a focus over a two-dimensional grid. Alice can send an image of the focus (I) or its Fourier transform (F) to Bob. Bob randomly switches between the two bases I and F. Only if the two bases are compatible (II or FF), the information encoded by Alice can be read out by Bob. In the two other cases (IF or FI), the information is low. Just like in BB84, a public channel is used for post processing including revealing the bases choices, detection of eavesdropping, error correction and privacy amplification.} \label{fig:scheme}
\end{center}
\end{figure*}

We implement a high-dimensional version of the BB84 protocol using the x,y spatial translation of single photons to encode information \cite{walborn2006quantum, tentrup2017transmitting}. A detailed description of our setup is given in the supplemental material. The working principle of the protocol is illustrated in Fig. \ref{fig:scheme}. We define detection areas on the two-dimensional plane representing the symbols of our alphabet. The detection areas span $10 \times 10$ pixels on our single-photon sensitive detector. All the areas are arranged in a two-dimensional grid of $32 \times 32$ symbols. In this way, we are able to encode $d=32^2=1024$ symbols in total, which allows a theoretical maximum of $I_\text{max}=10$~bit encoded in a single photon.
The protocol requires a second, mutually unbiased, basis to guarantee that a measurement in the wrong basis yields no information. In general it is always possible to use a Fourier transform to form this second basis. In optics, a single lens performs this task. Therefore, Alice and Bob both switch between an imaging path and a Fourier path. Only two of the four possible combinations will reveal all the information that Alice encoded to Bob. The two remaining cases will not provide any information.

\section{\label{sec:Results}Results}

\begin{figure}[htb]
\begin{center}
\includegraphics[width=\columnwidth]{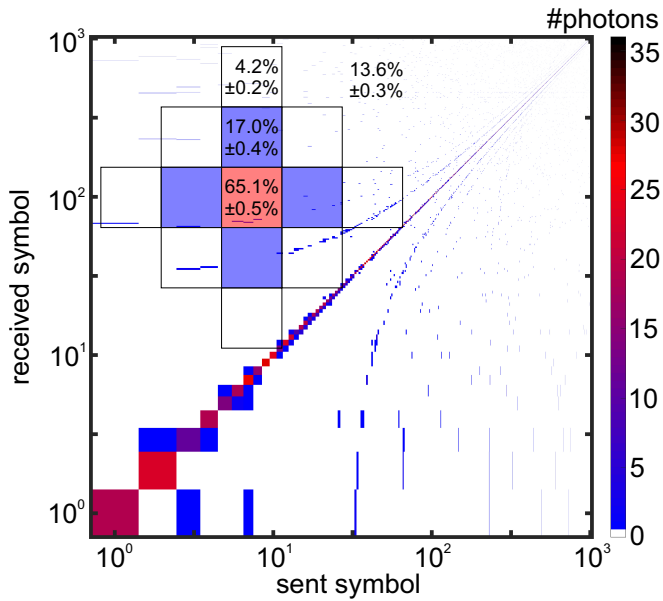}
\caption{Correlation map of the received symbol index versus the sent symbol index on a logarithmic scale in II configuration. The diagonal line indicates a strong correlation between the sent and the received symbols. In the top left corner, the hit distribution to the nearest and next-nearest neighbor symbols is shown. The events are visible in the correlation map as the lines shifted by the number of columns of the map. The average hit probability of the target area (red) is $P_0$, that of the four nearest neighbors (blue) is $P_1$, that of  the eight next nearest neighbors (white) $P_2$. The remaining probability is summed in $P_\text{rest}=1-P_0-P_1-P_2$. } \label{fig:II_map}
\end{center}
\end{figure}

First, we characterize the information content of the transmission from Alice to Bob. For this purpose, we analyze the two compatible bases choices of Alice and Bob (II and FF). Alice sends each symbol $x$ out of her alphabet $X$ individually, while Bob receives the symbol $y$ out of the alphabet $Y$. Per symbol $1000$ images are recorded on Bob's side. This step is performed for both compatible bases. In Fig. \ref{fig:II_map} the number of photons detected per symbol is shown in a log-log plot. In this figure, the joint probability function $p(x,y)$ is sampled, where $x$ is an element of the sent alphabet $X$ and $y$ from the received alphabet $Y$. We quantify the shared information between Alice and Bob by the mutual information \cite{nielsen2002quantum} 
\begin{equation}
I(X;Y)=\sum_{x \in X,\:y\in Y} p(x,y) \log_2 \left(\frac{p(x,y)}{p(x)p(y)} \right),
\label{eq:mutualinformation}
\end{equation}  
where $p(y)$ is the probability to measure symbol $y$ and $p(x)$ the probability of a sent symbol $x$. The maximum information Alice can send per symbol is $I(\text{Alice})= 10$~bit. Due to noise in the channel and in the detection and imperfections in the information encoding, the shared information between Alice and Bob is smaller. For the II and FF basis configuration, we calculated the sampled mutual information to be $I(X:Y)_\text{II}=8.3$~bit and $I(X:Y)_\text{FF}=8.1$~bit, respectively. The two main contributions to the noise are the cross talk to the neighboring detection areas, which was $13.3\%$ and the dark counts of the detector which was $13.8\%$.

Despite considerable experimental efforts, the probabilities used in the calculation of the mutual information are under-sampled with an average of $73$ detection events per symbol. This means that neighboring pixel cross-talk events are not accurately sampled, a problem that gets increasingly severe for larger alphabets. If Eve uses an optimal cloner \cite{bruss1999optimal}, the minimum fidelity for cloning-based individual attacks is $51.6 \%$ \cite{cerf2002security}. Introducing the average symbol hit probability $F$, the mutual information in equation \eqref{eq:mutualinformation} can be simplified to
\begin{align}
I(F)&=\log_2(d)+F \log_2(F) \nonumber\\ 
&+ (1-F) \log_2 \left( \frac{1-F}{d-1} \right), \label{eq:MIF}
\end{align}
where $d$ is the dimensionality of the basis. In our experiment, $F=68.7\%$. Since a large portion of the photons hits the neighboring areas, equation \eqref{eq:MIF} is an underestimate and can be refined by adding the hit probabilities $P_0$, $P_1$, $P_2$ and $P_\text{rest}$ defined in the top left corner of Fig. \ref{fig:II_map}. We assume the values $P_0$, $P_1$, $P_2$ and $P_\text{rest}$ are equal for each symbol and derive
\begin{align}
I_\text{AB}&=\log_2(d)\nonumber\\ 
&+P_0\log_2(P_0)+P_1 \log_2 \left(\frac{P_1}{4}\right)\nonumber\\
&+ P_2 \log_2 \left(\frac{P_2}{8}\right) + P_\text{rest} \log_2\left(\frac{P_\text{rest}}{d-13}\right).\label{eq:MInn}
\end{align} 
The resulting mutual information is $6.75\pm0.08$~bit in the II configuration and $7.03\pm0.04$~bit in the FF configuration.

\section{Discussion}
\begin{figure}[htb]
\begin{center}
\includegraphics[width=\columnwidth]{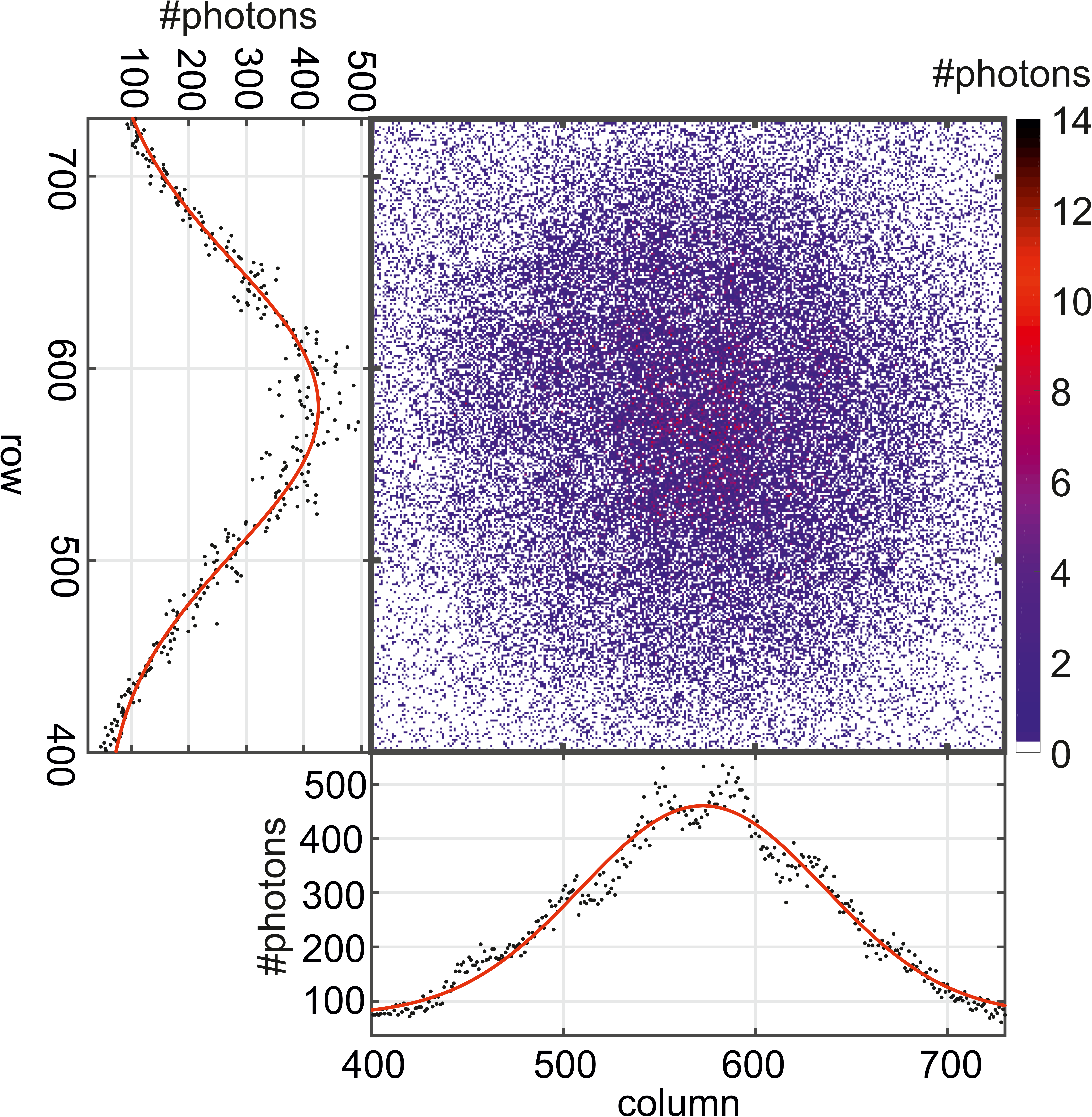}
\caption{Image integrated over all Bob's $1024$ different spot positions in the IF configuration. Projections of the signal are shown left of and below the plot by summing over the columns (lower panel) and rows (left panel). The photon counts follow a Gaussian spatial distribution as is evident from the red fit curves. }\label{fig:IF}
\end{center}
\end{figure}

We use Gaussian optics in our setup. As a result, we have Gaussian foci with finite width in the focus plane. The width results in crosstalk to the neighboring symbols, which reduces the mutual information as seen in equation \eqref{eq:MInn}. This mutual information results from averaging over the whole alphabet and is a lower bound for the security analysis. One important criterion for the security of QKD is that the basis choice of Alice remains hidden from Eve. In quantum mechanics orthogonal states can be distinguished. Therefore, if the attacker knows which of the two mutual unbiased bases is used, he can read out and resend the symbol without introducing errors. The Fourier transform of a Gaussian function is another Gaussian function, as seen in Fig. \ref{fig:IF}. In the Fourier basis, the probability to detect a photon is higher in the center than at the edges. If Alice sends all symbols of her alphabet with the same probability, Eve could therefore make a reasonable guess which basis is used. A photon detection at the edge of the detector is more likely to have been sent in the imaging basis, while a detection in the center is more likely in the Fourier basis. We measured the photon hit distribution for the two incompatible bases choices IF and FI with the same parameters as in the compatible case. In Fig. \ref{fig:IF}, the distribution is shown with a Gaussian fit.
The width in the columns is $89.9\pm1.7$ pixel and $106.7\pm1.9$ pixel in the rows together with $96.3\pm2.5$ pixel and $102\pm3$ pixel in the FI configuration.
To close the leak, Alice can adjust her send probability $p(k)$ to this Gaussian distribution. As a result, the information sent by Alice $I(\text{Alice})=-\sum^{d-1}_{k=0}p(k) \log_2(p(k))$ reduces from $10$~bit to $I(\text{Alice})_\text{II}=9.4$~bit and $I(\text{Alice})_\text{FF}=9.4$~bit. Consequently, the sampled mutual information with the hidden basis drops to \cite{walborn2006quantum}
\begin{align}
I_\text{hb}&=I(\text{Alice})+\sum^{d-1}_{k=0}p(k)F_\text{eff} \log_2(F_\text{eff}) \nonumber \\
&+ \sum^{d-1}_{k=0}\sum^{d-1}_{j=0,j\neq k} \frac{p(k)(1-F_\text{eff})p(j)}{1-p(k)} \log_2\left(\frac{(1-F_\text{eff})p(j)}{1-p(k)}\right)
\end{align}
with the effective fidelity $F_\text{eff}$ defined by $I(F_\text{eff})=I_\text{AB}$ in combining equation \eqref{eq:MIF} and \eqref{eq:MInn}. This results in $(F_\text{eff})_\text{II}=75.5\%$ and $(F_\text{eff})_\text{FF}=77.9\%$ leading to $(I_\text{hb})_\text{II}=6.3$~bit and $(I_\text{hb})_\text{FF}=6.6$~bit.         

In a postprocessing step via the public channel, Alice and Bob reveal their basis choices. They only keep the measurement results if they measured in two compatible bases, which bisects the key length. To check for eavesdropping, the quantum bit error rate of this sifted key needs to be calculated. We used the Gray code \cite{gray1953pulse} to encode the x and y position of the symbol in a bit string. In this way we reduce the bit error rate, since $31.3\%$ of the error is due to crosstalk to neighboring symbols. In the Gray code, neighboring symbols have a Hamming distance of only $1$. We calculated the averaged quantum bit error rate over all symbols to be $Q_\text{II}=7.8\%$ for the II configuration and $Q_\text{FF}=7.4\%$ for the FF configuration. We calculated the secret fraction of the key in case of intercept-resend attacks and infinite key length in the supplemental material.


\begin{figure}[H]
\begin{center}
\includegraphics[width=\columnwidth]{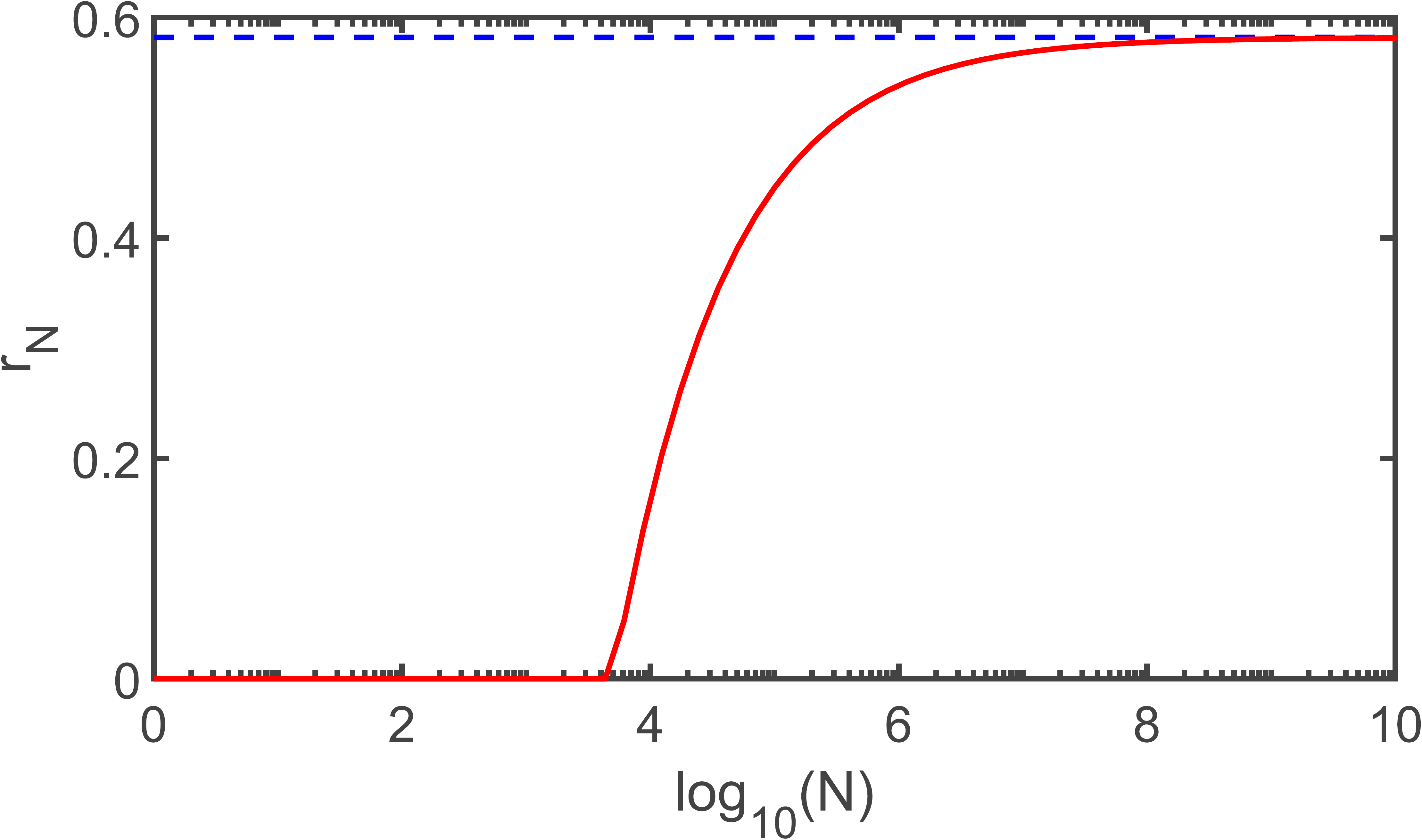}
\caption{The lower bound of the secret-key rate $r_N$per detected photon as a function of the logarithm of the key length $N$ (red). The blue dashed line represents the asymptotic limit of infinite key length. The failure probabilities are $\epsilon_\text{EC}=10^{-5}$ and $\epsilon_\text{PA}=2^{-15}$. The quantum bit error rate is $Q=0.08$.}\label{fig:rN}
\end{center}
\end{figure}

In order to analyse security arguments against collective attacks, we used finite-key considerations given in \cite{sheridan2010security,scarani2008quantum,cai2009finite}. In the case of a finite key length, $N < \infty$, failure probabilities in each step of postprocessing need to be considered. After sifting the key and removing the incompatible basis choices of Alice and Bob, the key length bisects. From this reduced key length, half the symbols are used to check for the presence of an eavesdropper. The next step is error correction to achieve an error-free key. Due to the finite key length the error correction has a finite failure probability and not all errors can be removed. Assuming a two-way cascade code \cite{brassard1994lecture}, this failure probability is $\epsilon_\text{EC} \sim 10^{-5}$ \cite{Martinez-Mateo:2015:DIR:2871401.2871407,tomamichel2014fundamental} in case of a $8\%$ bit error rate. To limit the maximum information of Eve, a privacy amplification step needs to be performed. With average bound privacy amplification \cite{bennett1995generalized,Ftentrgilbert2001privacy}, the information of Eve can be bound to $3 \times 10^{-10}$~bit with a failure probability of $\epsilon_\text{PA} = 2^{-15}$. In this case, the lower bound for the secret-key rate per photon of a $\epsilon=10^{-5}$ secure key is given by \cite{sheridan2010security,scarani2008quantum,cai2009finite} 
\begin{equation}
r_N=\frac{n}{N} \left(I_\text{AB} - I(\text{Eve}) -\frac{1}{n} \log_2(\frac{2}{\epsilon_\text{EC}})- \frac{2}{n}\log_2(\frac{1}{\epsilon_\text{PA}})\right). \label{eq:rate}
\end{equation}
 We neglect the failure probability introduced by smoothening the entropies. If both bases are used with equal probability, $n=0.25 N$ symbols can be used to create a key while $m=0.25 N$ symbols are used for parameter estimation to detect the presence of an eavesdropper. $I_\text{AB}$ is defined in equation \eqref{eq:MInn} and is the mutual information between Alice and Bob and 
\begin{align}
I(\text{Eve})&= -(1-F+\Delta F/\sqrt{m}) \log_2 (\frac{1-F+\Delta F/\sqrt{m}}{d-1}) \nonumber \\
&-(F-\Delta F/\sqrt{m})\log_2 (F-\Delta F/\sqrt{m})
\end{align}
is Eve's information assuming all channel errors are attributed to the presence of an eavesdropper \cite{cerf2002security}. We assume the worst-case values in parameter estimation for the fidelity $F$ by taking the standard deviation $\Delta F$ of the measured fidelity into account. This uncertainty in the fidelity is reduced by taking larger samples $m$ for parameter estimation. The remaining terms in equation \eqref{eq:rate} are the influence of the failure probabilities on the secret-key rate.

Figure \ref{fig:rN} shows the minimum secret-key rate as a function of the number of symbols. With increasing key length, the secret-key rate approaches its asymptotic limit, which is the difference between the shared information between Alice and Bob and the information of Eve. As seen in the figure, we can establish a non-zero secret-key rate starting from a key length of $5 \cdot 10^{3}$ symbols. Assuming an SLM with a maximum frame rate of $60$ fps, such a key can be generated in $\approx 3$ minutes. The secure key rate per photon asymptotically approaches $0.58$~ bit per photon. With the overall losses throughout the setup averaged over the four possible bases of $18.2\%$ and a quantum efficiency of our ICCD detector of $28\%$, we end up with a final secure key rate of $8$~bit per second. This rate can be improved by replacing the SLM in our setup by galvo mirrors. With an ICCD with $5000$~fps the final key rate can go up to $660$~bit per second.  

In principle, there could be a security loophole caused by the limited measurement range of the detection system, which is in our case the finite aperture of the ICCD \cite{bourassa2018entropic}. However, with the SLM we have full control of the prepared wavefronts and can therefore avoid that the light falls outside the detector. For the Fourier-transformed light, straightforward additional spatial filtering can be applied by Alice to not overfill Bob's detector and avoid this loophole. 

\section{Conclusion}
In this paper, we experimentally demonstrate high-dimensional QKD using spatially encoded photons. We encode an alphabet of $1024$ symbols and achieve a channel capacity of $7$~bit per detected photon. We discuss a solution to hide Alice's basis choice from Eve. Taking error correction and privacy amplification into account for finite key length, we show a secret-key fraction of $0.5$~bit per photon. For longer-distance communication, the combination of this work with multimode fibers \cite{amitonova2018multimode} appears attractive. 

\section*{Acknowledgments}
We would like to thank the Nederlandse Organisatie voor Wetenschappelijk Onderzoek (NWO) for funding this research. We thank Lyuba Amitonova, Jelmer Renema, Ravitej Uppu and Willem Vos for support and discussions. We also like to thank Valerio Scarani for giving us useful input for the finite-key formalism.   

\newpage

\bibliographystyle{apsrev4-1}
\bibliography{Bibliography}

\clearpage

\section{Supplemental Material}
\subsection{\label{sec:Setup}Setup}

\begin{figure*}[t!]
\begin{center}
\includegraphics[width=0.75\textwidth]{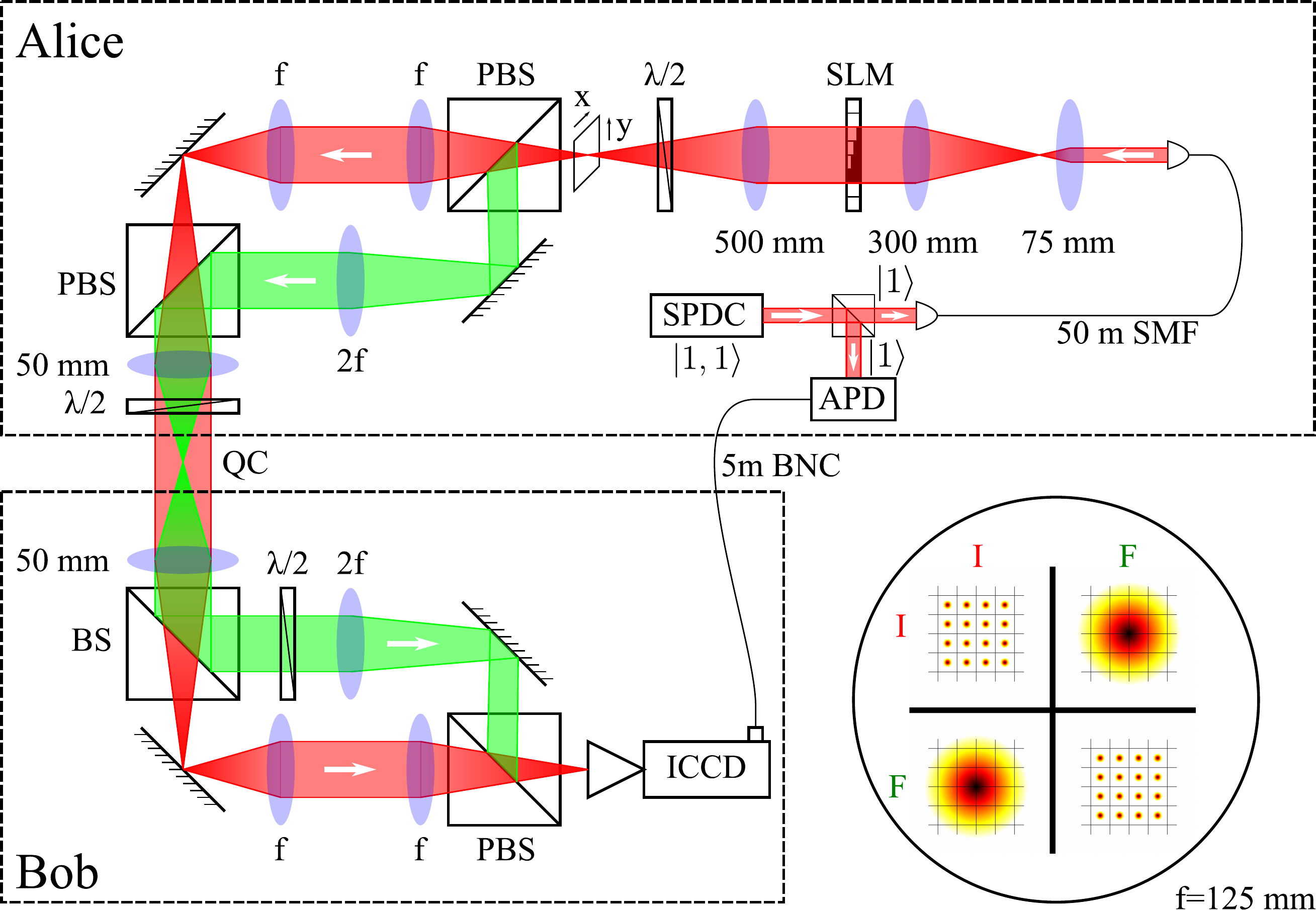}
\caption{Schematic representation of the setup. We generate photon pairs at $790$~nm by spontaneous parametric down-conversion (SPDC). One of the photons is coupled into a single-mode fiber (SMF) and the other is sent to an avalanche photodiode (APD) and used to trigger the camera. Information is encoded into the signal photon by translating the x and y position of the focus with a spatial light modulator (SLM) and a $500$~mm lens. Alice chooses between two paths with a half-wave plate and a polarizing beam splitter (PBS), one (green) with a single lens (2f) and one (red) with two lenses (f). After the two beams are merged again by a second PBS and the polarization information is erased by a second half-wave plate, the light is guided through the quantum channel (QC) with two $50$~mm lenses. Bob has the same set of lenses as Alice. His two paths are chosen randomly by a beam splitter (BS). The additional half-wave plate makes sure that all the light is directed to the camera (ICCD) by the last PBS.} \label{fig:setup}
\end{center}
\end{figure*}

Our setup consists of a single-photon source, a spatial encoder, optics, the free-space quantum channel between Alice and Bob and finally a decoder together with a single-photon-sensitive position-reading detector on Bob's side. The setup is schematically shown in Fig. \ref{fig:setup}. We use spontaneous parametric down-conversion (SPDC) \cite{tentrup2017transmitting,wolterink2016programmable} as a source of photon pairs, here called signal and herald. The wavelength of the generated photons is $790$~nm with a heralded single-photon count rate of $280$~kHz and a coincidence probability of $10\%$ measured with two avalanche photodiodes (APDs). The low single-photon count rate ensures a probability of less than $0.1\%$ to have more than one photon pair per pulse. The photon in the herald arm is detected by one of the APDs and signals a successful photon pair creation. The signal photon is sent through a $77$~m single-mode fiber (SMF) to add an optical delay. The fiber output coupler (Thorlabs F220FC-780) and a $75$~mm lens together with a $300$~mm lens expand the light to a collimated beam of $8.4$~mm FWHM to match the size of the spatial light modulator (SLM). We use a phase-only liquid-crystal SLM to write a blazed grating in the phase of the wavefront. A $500$~mm lens focuses the light to distinct x,y positions in its focal plane. With a half-wave plate and a polarizing beam splitter (PBS), Alice can switch between the two mutually unbiased bases. One basis is designed for imaging the light in a 4f configuration with two $125$~mm lenses. The other one performs a Fourier transform with a single lens with twice the focal length. Since the basis choice needs to be hidden from an eavesdropper, a second half-wave plate is put after Alice's last PBS to counteract the polarization rotation of the first half-wave plate. After being transmitted via the quantum channel with two lenses ($f=50$~mm), the photons are randomly split by a beam splitter (BS) and are again guided through an imaging or a Fourier transform path for decoding. The half-wave plate in the Fourier path ensures that all the light is reflected to the intensified charged-coupled detector (ICCD). The ICCD is triggered via a $5$~m BNC cable to match the detection window of the camera to the arrival of the signal photon.          

\subsection{\label{sec:Detection}Detection}

To detect the photons in a two-dimensional grid, we use an ICCD (Lambert HICAM 500S). It consists of an intensifier stage fiber-coupled to a CMOS camera of $1280 \times 1024$ pixels. The photocathode of the ICCD acts as a gate and is triggered by the herald photons at $280$~kHz. The delay between the trigger signal and the signal photon was measured to be $91$~ns. The gate width of the intensifier is $5$~ns. The CMOS camera is read out with $500$ frames per second. The variance of the read-out noise of the CMOS is 0.4 counts and a threshold of $5$ counts is set to filter the readout noise from the data. Moreover, a threshold on the size and intensity of detection events is set to between $2$ and $10$ pixels and between $1$ and $60$ counts, respectively, to remove unwanted spurious ion events. 

\subsection{Trojan-horse attacks}

The quantum channel connecting Alice and Bob could be used as a door by an eavesdropper to read out the state of Alice's and Bob's devices, making the setup vulnerable against Trojan-horse attacks \cite{gisin2006trojan}. To counteract such attacks, the optical devices should only be active when the photons are sent. In our setup that could be realized by replacing the mechanical switch of the half-wave plate with a fast electro-optic modulator and the liquid-crystal SLM by a faster digital micromirror device. They could be synchronized to the photon arrival. Another countermeasure is to use bandpass filters and optical isolators at the entrance of Alice's device. Alice should also use auxiliary detectors to detect any light entering her device to detect attacks. In the finite-key regime the security of leaky decoy-state BB84 has been investigated \cite{wang2018finite}.

\subsection{Intercept-resend attack}
 In this attack, Eve intercepts a fraction $\eta$ of the quantum states and performs projective measurements randomly choosing one of the two mutually unbiased bases. Eve resends her measurement result, introducing an error due to the collapse of the wavefunction. For the security of the protocol, we need to ensure that the information Eve can gain from intercepting the communication is lower than the mutual information between sender and receiver \cite{gisin2002quantum,scarani2009security}. Assuming an intercept-resend attack, the information Eve can learn is $I(\text{Eve})=\frac{\eta}{2}I(\text{Alice})$ with $\eta$ the fraction of intercepted photons and $I(\text{Alice})$ the sent information of Alice. Averaged over the compatible bases, we find $I(\text{Alice})=9.4$ bit. An eavesdropper can extract a maximum of $I(\text{Eve})=\eta 5.72$ bit. Therefore, just as in the case of the original BB84, information gain is only possible at the expense of disturbing the signal \cite{nielsen2002quantum}. An eavesdropper will be recognized in postprocessing, since he collapses the wavefunction and thereby increases the error rate of the key generated by Alice and Bob. Alice and Bob will have to compare a random part of their key to decide if they have been eavesdropped. Intercepting a fraction of $\eta$ photons, an attacker introduces an error of 
\begin{equation}
E_\text{Eve}=\frac{\eta}{2}\frac{d-1}{d} ,
\end{equation}
where $d$ is the number of symbols. To calculate the quantum bit error rate of the sifted key, we used the Gray code \cite{gray1953pulse} to encode the x and y position of the symbol in a bit string. In this way, we reduce the bit error rate, since $31.3\%$ of the error is due to crosstalk to neighboring symbols. In the Gray code, neighboring symbols have a minimum hamming distance of $1$. We calculated the averaged quantum bit error rate over all symbols to be $Q_\text{II}=0.078$ with a standard deviation of $\Delta Q_\text{II}=0.042$ for II configuration and $Q_\text{FF}=0.074$ and $\Delta Q_\text{FF}=0.029$ for FF configuration. We assume Alice and Bob set their threshold to detect eavesdropping to a bit error rate of $Q+\sigma$, where $Q$ is the averaged quantum bit error rate. In this case Eve could only intercept a fraction $\eta$ of the photons. 

\subsection{Basis guess fidelity}

\begin{figure}[htb]
\begin{center}
\includegraphics[width=\columnwidth]{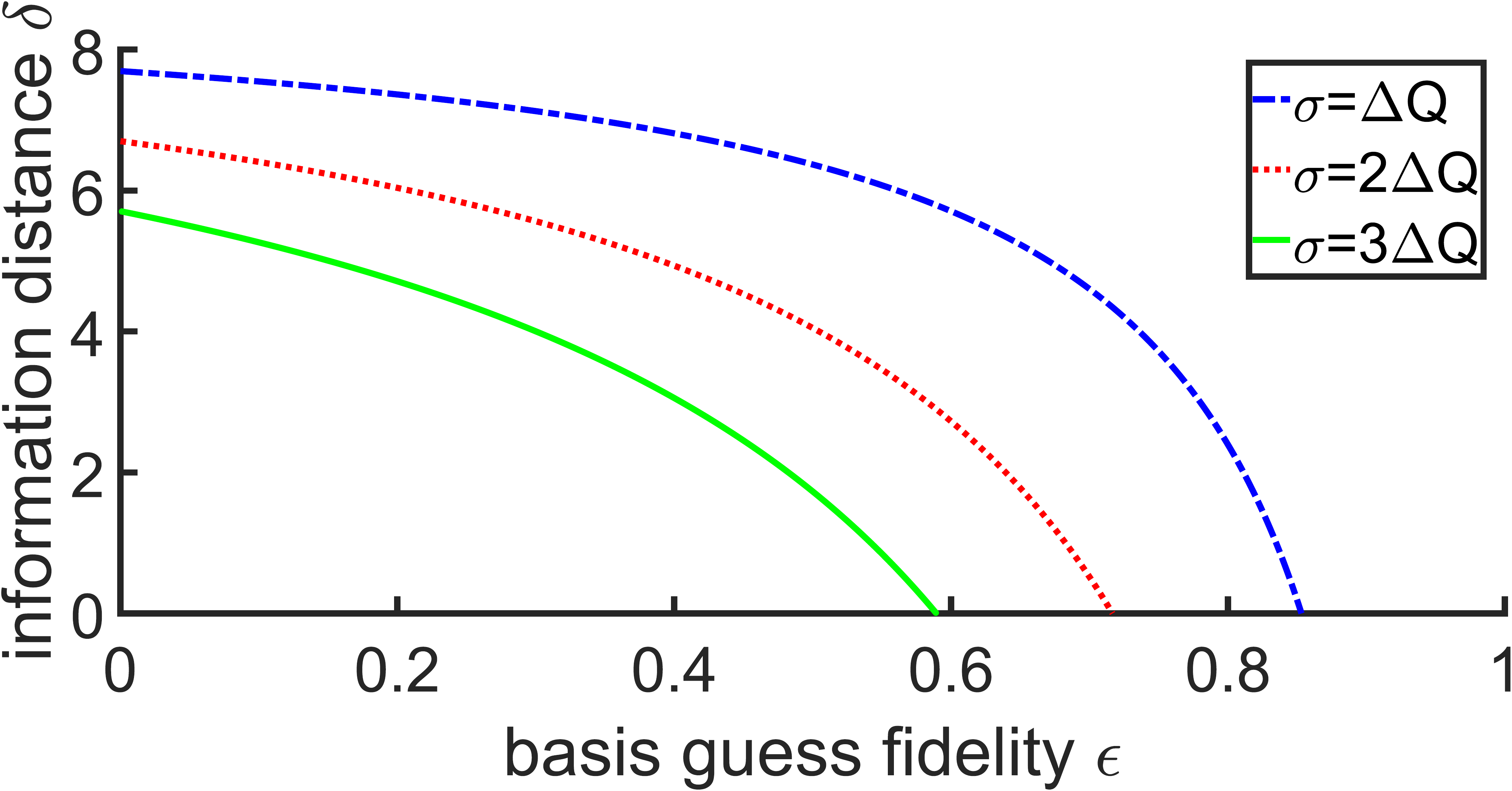}
\caption{The minimum secret information distance $\delta$ against the basis guess fidelity $\epsilon$ for three different thresholds $\sigma$.}\label{fig:delta}
\end{center}
\end{figure}

In practical QKD, Alice's basis choice could leak to an eavesdropper via side channels or imperfect encoding. To include this into the model, we added a guess fidelity of $\epsilon$. Eve can not guess the basis if $\epsilon=0$, while $\epsilon=1$ means that Eve knows Alice's basis choice. In our experiment, we measured $\epsilon\sim 0.15$ by a correlation measurement performed with classical light. Eve can then extract the information
\begin{equation}
I(\text{Eve})=\frac{\eta}{2}\left(1+\epsilon \right)I(\text{Alice})\label{eq:IEve}
\end{equation}
from what Alice sends. Thereby she adds an additional error of
\begin{equation}
 Q_\text{Eve}=\frac{1}{2}\left(1-\epsilon \right)\frac{d-1}{d}.
\end{equation}
To detect eavesdropping, Alice and Bob must set an error threshold $\sigma$. The error rate introduced by Eve's perturbation of the quantum channel has to be lower than this threshold to stay unnoticed. The quantum bit error rate including an eavesdropper is
\begin{align}
Q_\text{Total}&=(1-\eta)Q +\eta \left(Q +(1-Q)Q_\text{Eve} \right)\\
&=Q +(1-Q)\eta Q_\text{Eve}.
\end{align}
From the relation
\begin{equation}
Q_\text{Total}\leq Q+\sigma,
\end{equation}
the maximum fraction of intercepted photons is
\begin{equation}
\eta_\text{max}=\frac{\sigma}{(1-Q)Q_\text{Eve}}, \label{eq:etamax}
\end{equation}
which depends on the fidelity to guess the correct basis $\epsilon$ and the threshold $\sigma$. The minimum fidelity between Alice and Bob reduces from $F$ to
\begin{equation}
F_\text{Total}=F\left(1-\eta Q_\text{Eve}\right).
\end{equation}
Now it is possible to calculate the distance of information between Bob and Eve, which is a measure for the secure key rate. The information distance is defined as
\begin{equation}
\delta=I_\text{AB}-I(\text{Eve}).
\end{equation}
The amount of information Bob receives depends on the amount of information Alice transmits and on the channel noise. Therefore 
\begin{equation}
I(\text{Bob})=Q_\text{Total}I(\text{Alice}). \label{eq:IBob}
\end{equation}
Combining equation \eqref{eq:IEve} and \eqref{eq:IBob}, the information distance can be written as
\begin{equation}
\delta=\left[(1-Q_\text{Total})-\frac{\eta}{2}(1+\epsilon) \right]I(\text{Alice}).
\end{equation}
By substituting $\eta_\text{max}$ from equation \eqref{eq:etamax} to this expression, the minimum information distance can be plotted as a function of $\epsilon$ and $\sigma$ in figure \ref{fig:delta}. Compared to equation \eqref{eq:rate} in the main article the prefactor $n/N$ does not appear, since already a small fraction of the key is enough for parameter estimation. Moreover, the error correction and parameter estimation as well as the uncertainties about Eve's entropies lower the secret fraction. 

The information distance $\delta$ grows monotonically with decreasing threshold $\sigma$, but becomes smaller with increasing $\epsilon$, as visible in Fig. \ref{fig:delta}. If Eve knows Alice's basis choice ($\epsilon=1$), her measurements will no longer add noise, which allows here to intercept the quantum communication without being detected. In comparison to the finite-key-length secret fraction in the case of collective attacks, the values the minimum secret information $\delta$ is larger.

\end{document}